\documentstyle[amssymb,preprint,aps]{revtex}
\begin{document}

\draft

\title{Dielectric susceptibility of the Coulomb-glass}
\author{
A. D\'\i az-S\'anchez,$^{1,3}$ M. Ortu\~no,$^{1}$ M. Pollak,$^{2}$
A. P\'erez-Garrido,$^{1}$ and A.\ M\"obius$^3$}
\address{
$^1$Departamento de F\'\i sica, Universidad de Murcia,
E-30071 Murcia, Spain,\\
$^2$Department of Physics, University of California at 
Riverside, Riverside, California 92521,\\
$^3$Institut f\"ur Festk\"orper- und Werkstofforschung,
D-01171 Dresden, Germany\\
}

\date{\today}
\maketitle

\begin{abstract}

We derive a microscopic expression for the dielectric susceptibility 
$\chi$ of a Coulomb glass, which corresponds to the definition used in 
classical electrodynamics, the derivative of the polarization with 
respect to the electric field. The fluctuation--dissipation theorem tells 
us that $\chi$ is a function of the thermal fluctuations of the dipole 
moment of the system. We calculate $\chi$ numerically for 
three--dimensional Coulomb glasses as a function of temperature 
and frequency. 
\end{abstract}

\section{Introduction}

The subject of this paper is the calculation of the dielectric
susceptibility in Coulomb glasses, a term used for Anderson insulators 
with Coulomb interactions between the localized electrons. We consider 
situations deep in the insulating phase, when quantum energies $t$ 
arising from tunneling are much smaller than the other important 
energies in the problem, i.e., Coulomb interactions and random 
energy fluctuations. The model also applies to systems in the quantum 
Hall regime far away from the peaks, for which the conductivity is 
exponentially small as compared to $e^2/h$ and the conduction mechanism 
is by variable range hopping between localized states. The model can be 
easily extended to granular metals in the insulating regime.
 
Previous calculations of the dielectric susceptibility of the Coulomb 
glass have used an expression directly obtained from the analogy between 
Coulomb and spin glasses \cite{DL84}. However, the non-local character 
of the processes involved in Coulomb glasses makes this expression of 
the dielectric susceptibility inappropriate. Furthermore, it does not 
corresponds to the standard definition of the dielectric susceptibility. 
The first aim of this work is to present a microscopic expression of the 
dielectric susceptibility $\chi$ valid for the Coulomb glass, and to 
apply the fluctuation--dissipation theorem to this expression. We also 
want to calculate $\chi$ at very low temperatures. To this end we have 
to take into account that interactions, and specially those of 
long--range character, drastically change the properties of systems with 
localized states\cite{PO85,SE84}. Most properties of these systems are 
affected by electron correlations, and such effects cannot be described
by one--particle densities of states or excitations. To deal with 
complex excitations, methods were developed 
\cite{PO97,MP96,MP91,ST93,TE93,DM97,DM98,MN97,MT97} to obtain the 
low lying states and energies of Coulomb glasses.

The paper is organized as follows: Section II introduces the Coulomb 
glass models used for the numerical calculations. Section III presents 
the derivation of a microscopic expression for the dielectric 
susceptibility of Coulomb glasses. Section IV describes how the 
low-energy many-particle states are obtained numerically and gives a 
method to calculate the dependence of the dielectric susceptibility 
with the frequency. In section V, we present the results obtained for 
the dielectric susceptibility of Coulomb glasses at very low 
temperatures and the dependence on temperature, and frequency.
Finally, in section VI we extract some conclusions.

\section{Models}

Our next results apply to any model of a Coulomb glass but to be 
definite we performed numerical simulations for the two most common 
models: the standard model with a uniform random potential 
distribution and the classical impurity band model (CIB).

Efros and Shklovskii proposed a practical model to represent Coulomb 
glass problems with localized electronic states, which has been widely 
used and extended \cite{PO85,SE84}. This model is represented by 
the standard tight-binding Hamiltonian:
\begin{equation}
H=\sum_i\phi_i n_i +\sum_{i<j} \frac{e^2}{4\pi\epsilon_0}
{\frac{(n_i-K)(n_j-K)}{r_{ij}}} \label{standm}
\end{equation}
where $n_i\in \{0,1\}$ denotes the occupation number of site $i$.
We use rationalized units, unlike in most works on the Coulomb glass, 
because in this problem they constitute the most convenient choice. 
We will consider sites at random positions, with a density $\rho$ equals 
to 1, and simulate the disorder by a ramdon potential in each site 
$\phi_i$, uniformly distributed between $-W/2$ and $W/2$. $r_{ij}$ 
is the distance between sites $i$ and $j$ according to periodic boundary 
conditions in the sense of \onlinecite{MR53} in the perpendicular 
directions to the applied electric field. Charge neutrality is 
achieved by a background compensation charge $-K$ at each lattice site.  

The classical impurity band model (CIB) is a realistic representation of 
a lightly doped semiconductor in which the random potential arises from 
the minority impurities \cite{ES85}. Here we consider an n-type, 
partially compensated semiconductor with donor concentration 
$N_{\rm D}$, and acceptor concentration $N_{\rm A}=KN_{\rm D}$. The 
Hamiltonian is given by:
\begin{equation}
H=\frac{e^2}{4\pi\epsilon_0} \left( \frac12\sum_{i\not=j} 
\frac{(1-n_i)(1-n_j)}{r_{ij}}-\sum_{i\nu} \frac{(1-n_i)}{r_{i\nu}}
\right)
\label{cibm}
\end{equation}
where the donor occupation number $n_i$ equals 1 for occupied donors,
and 0 for ionized donors. The index $\nu$ runs over the acceptors and
$r_{i\nu}=\mid {\bf r}_i-{\bf r}_\nu \mid$, with ${\bf r}_i$ being the
donor coordinates and ${\bf r}_\nu$ the acceptor coordinates. We chose 
the donor density $\rho$ equal to 1 and imposed again periodic boundary 
conditions. For numerical reasons we constrained the nearest neighbor 
distance to be larger than 0.5. We used K=0.5 because the interaction 
effects are largest there.

\section{Susceptibility of the Coulomb glass}

We now proceed to obtain a proper microscopic expression for the 
dielectric susceptibility of Coulomb glasses. It is applicable to an 
arbitrary three--dimensional model of Coulomb glass as long as the 
interaction between charges goes as $1/r$.

A certain analogy between the spin glass and the Coulomb glass lead to 
an incorrect expression for the dielectric susceptibility of the Coulomb 
glass. In some sense the local spin $s_i$ in the spin glass is analogous 
to the site occupation $n_i$ in the Coulomb glass. If only $n_i=0$ and 
$n_i=1$ are allowed due to strong on-site interaction, then the analogy 
is with spins $s_i=1/2$. The magnetic field in the spin glass is then 
analogous to the chemical potential in the Coulomb glass. However, this 
analogy should not be pushed too far and does not apply to 
polarizabilities. The magnetic polarizability is the change of total 
spin induced by the magnetic field, but the electric polarizability is 
the change in the electric polarization induced by an electric field, 
not the change in the total occupation number induced by a change of a 
global potential as the direct analogy would have it. The basic 
difference between the two susceptibilities can be understood more 
clearly by realizing that the magnetic polarization 
$\langle s_i\rangle_T$ comes from field-induced flips of spatially fixed 
spins, where $\langle ... \rangle _T$ refers to thermal average. An 
analogous electric polarization can come from flips of local dipoles, 
but in many systems it involves a field induced displacement of charges 
(there is no magnetic equivalent to this because there are no magnetic 
charges). Such a polarizability is thus not represented by 
$\langle n_i \rangle_T$ responding to a potential (as in 
\onlinecite{DL84} and \onlinecite{GY93}) but by 
$\sum_i x_i \langle n_i \rangle_T$ responding to a field (as in 
\onlinecite{PG61}). The dielectric susceptibility $\chi$ then is:
 \begin{equation}
\chi=\frac{1}{\epsilon_0}{\partial P\over \partial E}
\end{equation}
where $E$ is the total electric field. 

Our first aim is to obtain a microscopic expression for Coulomb glasses 
of the classical definition of the dielectric susceptibility. Let us 
assume that we apply an electric displacement $D$. This will 
induce a polarization $P$ equal to 
\begin{equation}
P=\frac eN\sum_ix_i\langle \Delta n_i\rangle _T 
\end{equation}
where $x_i$ is the position of site $i$, $N$ is the number of sites, and 
$\langle\Delta n_i\rangle _T$ is the change in the average occupation of 
site $i$ due to the applied electric displacement. In the linear 
approximation, the change in the average occupation of site $i$ due to a 
general change in the potential is given by: 
\begin{equation}
\langle \Delta n_i\rangle _T=\sum_j\frac{\partial \langle n_i\rangle _T}
{\partial \phi _j}\Delta \phi _j 
\end{equation}
where $\Delta\phi_j$ is the change in potential at site $j$. The partial 
derivative appearing in this expression is proportional to the local 
susceptibility $\chi_{ij}$:
\begin{equation}
\frac{\partial \langle n_i\rangle _T}{\partial \phi _j}=\frac{e\epsilon_0}
{T}\chi _{ij}
\end{equation}
where $T$ is the temperature, and the 
Boltzmann constant $k_B$ is taken to be 1 throughout the paper.

The change in potential corresponding to a uniform electric displacement 
is $\Delta \phi_i = D x_i/\epsilon_0$, so the ratio between $P$ and $D$ 
is:
\begin{equation}
\frac{\partial P}{\partial D}=\frac {e^2}{T N}\sum_{ij}x_i\chi_{ij} x_j
\equiv \chi_0.
\label{local}
\end{equation}

To calculate the dielectric susceptibility numerically it is 
convenient to apply the fluctuation-dissipation theorem to $\chi_0$ in 
order to rewrite it in terms of  thermal fluctuations of the dipole 
moment. Taking into account the expression for the thermal average of
the site occupation, $\langle n_i\rangle_T$, it is easy to obtain that 
its derivative with respect to the potential in $j$, i.e., the local 
susceptibility $\chi_{ij}$, is equal to the fluctuation in the electron 
occupancy of the two sites involved:
\begin{equation}
\chi_{ij}= \langle n_in_j\rangle _T-
\langle n_i\rangle _T \langle n_j\rangle _T
\end{equation}
Using this equation in expression (\ref{local}) for $\chi_0$, we 
arrive at
\begin{eqnarray}
\chi_0 &=& \frac {e^2}{T N}\sum_{ij}x_i(\langle n_in_j\rangle
_T-\langle n_i\rangle _T\langle n_j\rangle _T)x_j \nonumber\\
&=&\frac {1}{T N} (\langle d^2\rangle _T-\langle d\rangle _T^2)
\label{dipp}
\end{eqnarray}
where $d=e\sum_{i}x_in_i$ is the dipolar moment of the sample. The 
dielectric susceptibility is a function of the thermal fluctuation of 
the dipolar moment. 

Our computer simulations can per force involve only systems of 
mesoscopic size. For the macroscopic susceptibility we can 
imagine building a macroscopic system of many mesoscopic cubes of 
linear size $L$ arranged to fill the space. Each of these samples 
corresponds to a particular realization of the random positions and
energies of the sites involved. The total electric field that a given 
microscopic sample feels is the sum of the applied field, $D$, and the
induced field. If the applied field is uniform, and the polarizabilities
of all the samples were the same, the polarization would also be
uniform and the induced field would come only from the boundary of the 
sample. We then have (in our units): 
\begin{equation}
\epsilon_0 E=D-P \label{total}
\end{equation}
and get from Eqs.\ (\ref{dipp}) and (\ref{total}):
\begin{equation}
\chi={\chi_0\over 1-\chi_0}.\label{fin}
\end{equation}
At very low frequencies most samples are conducting and we have an 
effectively uniform distribution of $\chi_0$ over the computer ``samples''.
At lower frequencies there is no mechanism mitigating the effect broad 
distribution of $\chi_0$ and the argument leading to Eq.\ (\ref{fin}) fails 
and the problem becomes Clausius-Mossotti-like. A general approach to this 
problem for random media has been given in \onlinecite{P71,PP77}. Here we 
shall avoid the inherent complications of such a computation and assume 
that Eq.\ (\ref{fin}) is approximately valid even at higher frequencies if 
we use for $\chi_0$ a  value averaged over many computer realizations.     
The relation $\sigma=i\omega\epsilon$ means that a finite dc conductivity 
implies an infinite dc dielectric susceptibility. A proper calculation of
the DC conductivity can be done by percolation in configuration space,
but it is a difficult problem requiring huge numerical efforts so that
for three-dimensional systems we could only consider very small samples.
Our approximate calculation of the divergence of the susceptibility
allows us to estimate the variation of the DC conductivity with
temperature as we will see.

\section{Numerical Procedure}

\subsection{Low-energy configurations}

We calculate the dielectric susceptibility at very low temperatures 
making use of the ground state and the very low-energy configurations of 
the systems. With the procedure that we briefly discuss bellow we obtain 
the first 5.000 many-particle configurations and calculate their dipole
fluctuations, Eq. (\ref{dipp}).

We find the low-energy many-particle configurations by means of a 
three-steps algorithm \cite{DM98}. This comprises local search 
\cite{PO97,MP96}, thermal cycling \cite{MN97}, and construction
of ``neighbouring'' states by local rearrangements of the charges 
\cite{PO97,MP96}. The efficiency of this algorithm is illustrated in 
Ref.\ \onlinecite{DM98}. In the first step we create an initial set 
${\cal S}$ of metastable states. We start from states chosen at random 
and relax these states by a local search algorithm which ensures 
stability with respect to excitations from one up to four sites. In the 
second step this set ${\cal S}$ is improved by means of the thermal 
cycling method, which combines the Metropolis and local search 
algorithms. The third step completes the set ${\cal S}$ by 
systematically investigating the surroundings of the states previously 
found. At the end we only keep configurations with a fix number of
electrons, so we work with canonical ensembles.

\subsection{Frequency dependence}

At finite frequencies only transitions with characteristic time
$\tau_{IJ}$ shorter than the inverse of the frequency contribute to 
the susceptibility. Thus, for a given frequency, we consider two 
configurations as connected if their $\tau_{IJ}$ is shorter than the 
inverse of the frequency, and we group the configurations in clusters 
according to these connections.

The characteristic transition time between configurations $I$ and $J$ is 
\cite{PO85},
\begin {equation}
\tau_{IJ}=\omega_0^{-1} \exp \left( 2\sum r_{ij}/a \right)
  \exp \left( E_{IJ}/T \right)/Z  
\end{equation}
In this equation, the quantity $\omega_0$ is a constant of the order of
the phonon frequency, $\omega_0\sim 10^{13}\ {\rm s^{-1}}$.  The sum is 
the minimized sum over all hopping distances between sites which change 
their occupation in the transition $I \rightarrow J$. $a$ denotes the
localization radius, $E_{IJ} = \max (E_I,E_J)$ where $E_I$ is the 
energy of the state $I$, and $Z$ is the partition function.

We calculate the susceptibility of each cluster through 
Eq.\ (\ref{dipp}), assuming thermal equilibrium in the cluster. 
The glassy nature of our systems is responsible for the existence
of the clusters, which indicate the non-ergodicity of the systems
for times shorter than the critical time connecting all the configurations
in a single cluster. Each realization of the systems will be in a given 
cluster and will not see the other clusters. The probability to be in
a cluster depends on the history of the system and is very difficult to 
estimate. In order to obtain averages of the susceptibility, we will
assume that the weight of each cluster is proportional to its partial 
partition function, which constitutes the simplest possible assumption. 
The results are finally averaged over many 
different disorder realizations.

\section{Results}

If we take into account all types of transitions, including the slowest 
ones, Coulomb glass behaves like a conductor and is able to screen 
fully as its susceptibility diverges. But small samples may not have 
excitations which carry electrons across the entire sample and produce 
nearly equipotential surfaces at the two opposite edges. So we must 
consider samples above a certain critical size at which the steady state 
susceptibility $\chi$ diverges. We found that this critical size is 
about 200 sites for both models (Eqs.\ (\ref{cibm}) and (\ref{standm})). 
Above this size the results are practically independent of size.

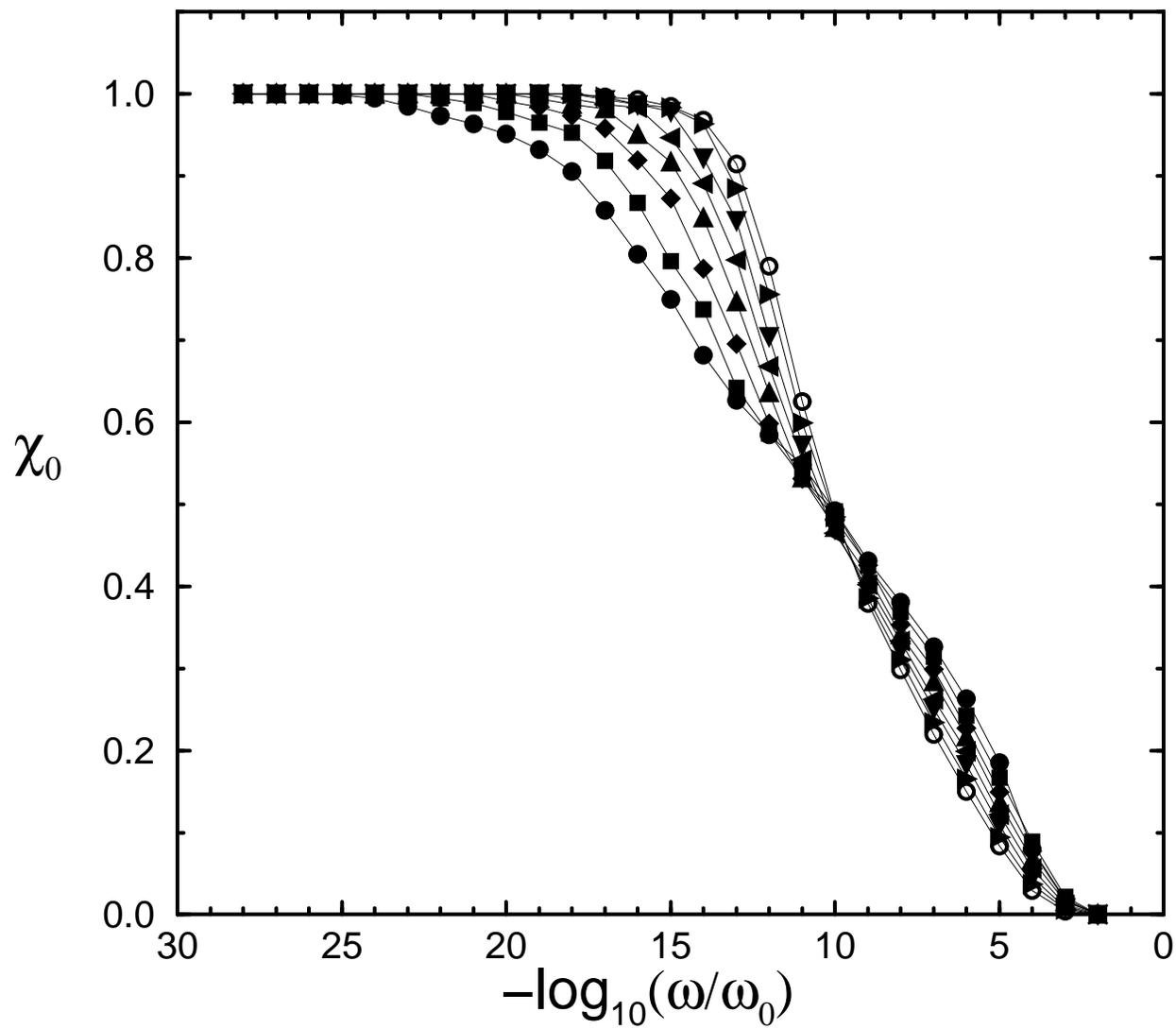
\begin{figure}
\caption{ Averaged values of $\chi_0$ obtained for the standard model, 
plotted against frequency, for several values of the temperature as
follows: 0.006 ($\bullet$), 0.008 ($\blacksquare$), 0.01 
($\blacklozenge$), 0.012 ($\blacktriangle$), 0.014 
($\blacktriangleleft$), 0.016 ($\blacktriangledown$), 0.018 
($\blacktriangleright$), and 0.02 ($\circ$).The disorder energy is 
$W=2$ and the localization radius is $a=0.2$.
}
\label{fig1}
\end{figure}

Fig.\ \ref{fig1} shows the average value of $\chi_0$ as a function of 
frequency for several temperatures. The plots are for the standard model 
of size $N=256$. The localization radius is $a=0.2$, which is maintained 
throughout the paper. At low frequencies $\chi_0$ increases with $T$ 
while at high frequencies it decreases with $T$. The reason is that at 
small $\omega$ hopping extends over many configurations, and the main 
effect of $T$ is to enhance the transition rates. At large $\omega$ 
hopping is between two optimal configurations for that frequency (or 
even two sites) and the main effect of $T$ is to equalize the occupation 
probabilities. This bears analogy with uncorrelated hopping conductivity 
which increases strongly with $T$ as $\omega \rightarrow 0$ and behaves 
as $1/T$ at high frequency.     

The result for the CIB model are very similar to those for the standard 
model and so we do not show them explicitly. The results for the CIB 
model roughly correspond to those for the standard model with an 
effective disorder energy of approximately 3. 

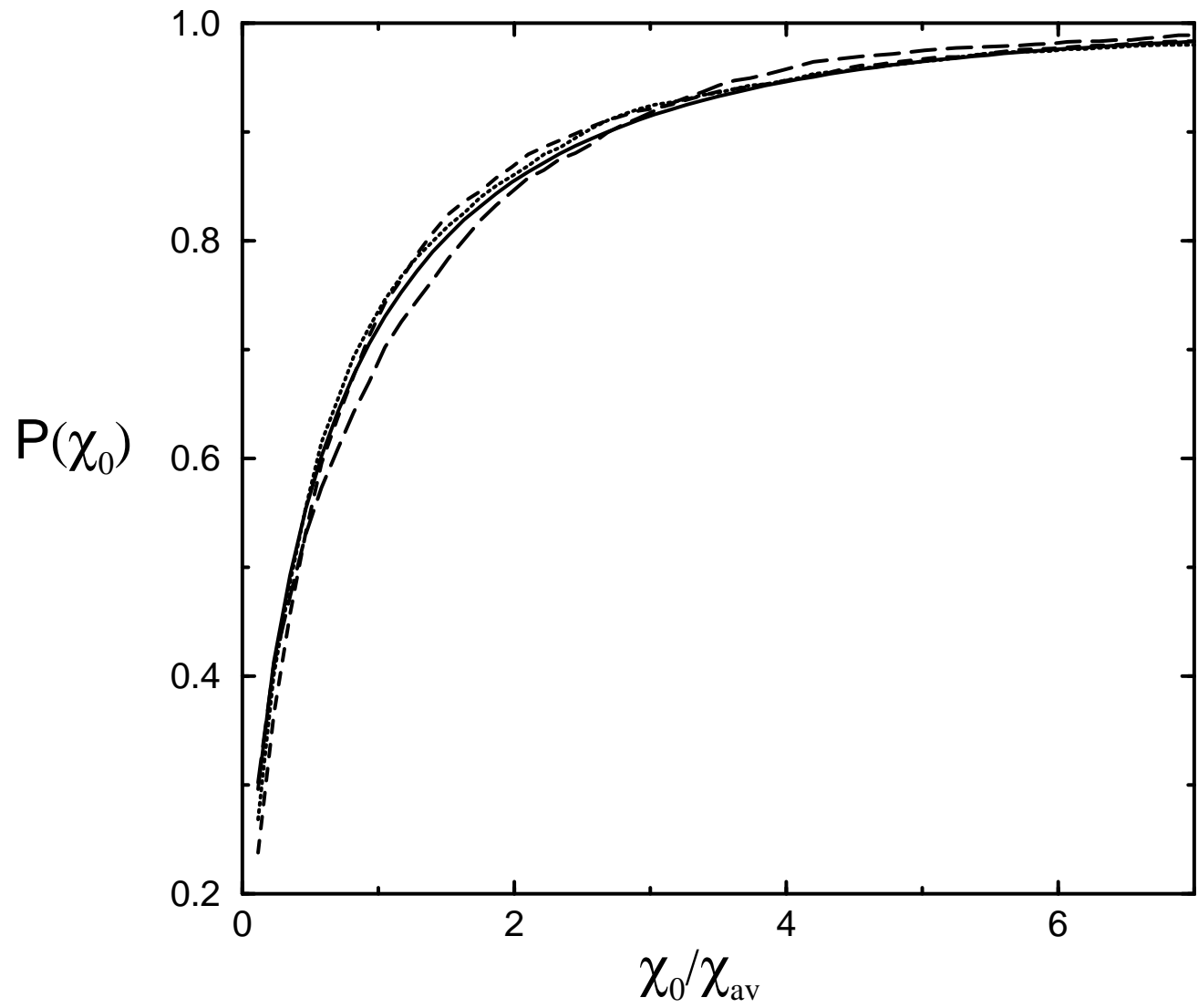
\begin{figure}
\caption{Accumulated distribution probability of $\chi_0$ 
for $N=256$ and $\omega \rightarrow 0$ (dotted curve), 
$\omega=10^{3}\ {\rm s^{-1}}$ (dashed curve), 
$\omega=10^{7}\ {\rm s^{-1}}$ (long dashed curve).The solid 
curve corresponds to the fit explained in the text.
$\chi_{\rm av}$ is the average value at each frequency.
}
\label{fig2}
\end{figure}

As already mentioned, an accurate calculation of the frequency dependent 
macroscopic susceptibility requires the distribution $f(\chi_0)$, not 
only the average value of $\chi_0$. Fig.\ \ref{fig2} shows the 
integrated distribution 
$P(\chi_0)=\int_{0}^{\chi_0} f(\chi_0')\ d\chi_0'$ for $N=256$ at 
$T=0.01$ for different values of the frequency, $\omega \rightarrow 0$ 
(dotted curve), $\omega=10^{3}\ {\rm s^{-1}}$ (dashed curve), 
$\omega=10^{7}\ {\rm s^{-1}}$ (long dashed curve). The solid curve is a 
plot of the function $1-\exp\{-(x\lambda)^{\alpha}\}$ with the 
parameters $\lambda=1.5$ and $\alpha=0.6$. In the range examined this 
form of integrated distribution fits (with varying $\lambda$ and 
$\alpha$) fairly well our data for all $T$ and $\omega$. The broad 
character of the distribution indicates large mesoscopic fluctuations 
and shows a need to examine in the future the accuracy of our 
approximation by taking proper account of the distribution of $\chi_0$. 

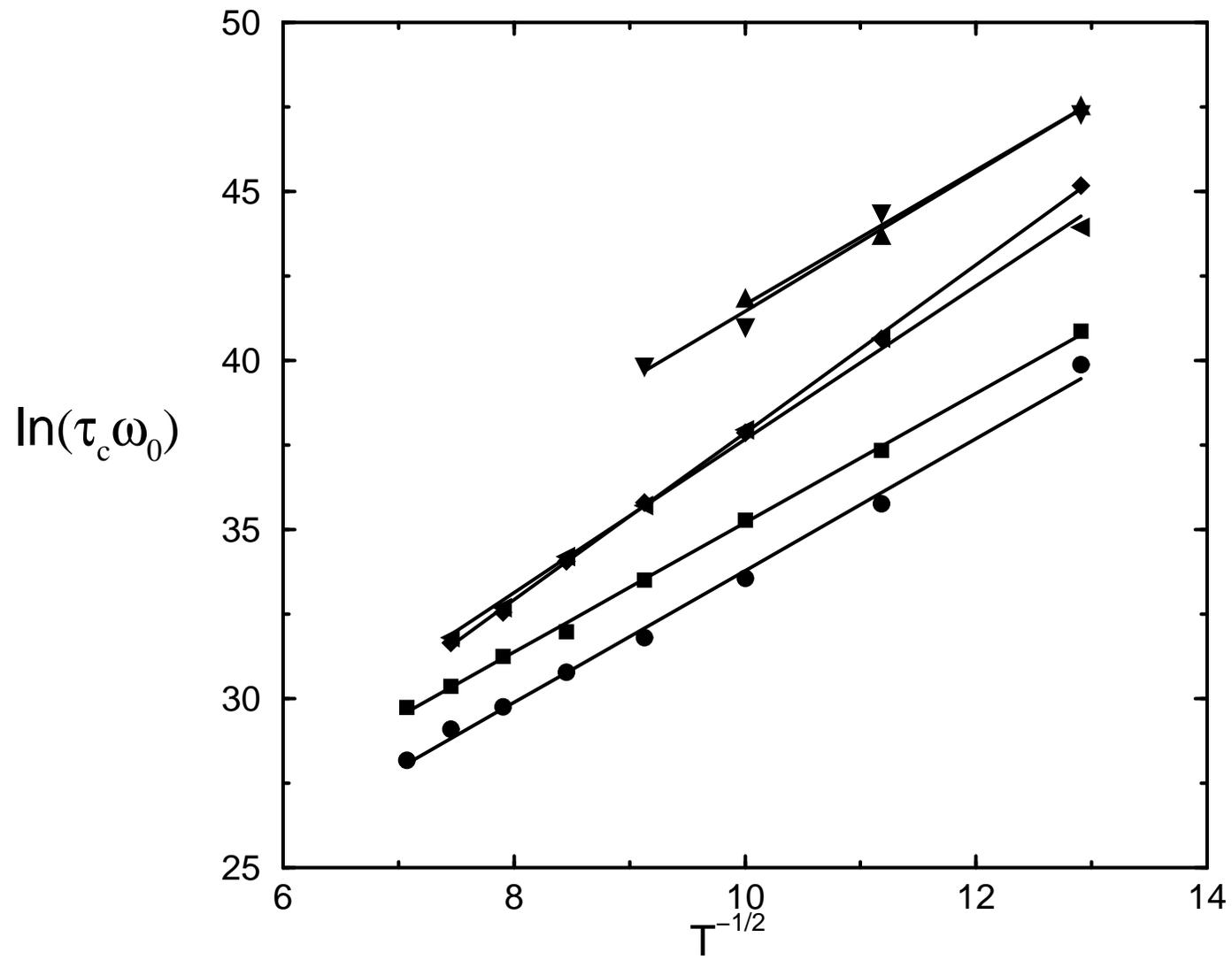
\begin{figure}
\caption{Logarithm of critical time $\tau_{\rm c}$ for reaching the 
static susceptibility as a function of $T^{-1/2}$ for four sizes of 
the standard model, $N=64$ ($\bullet$), 128 ($\blacksquare$), 216 
($\blacklozenge$), 512 ($\blacktriangle$), and for two sizes of the 
CIB model, $N=216$ ($\blacktriangleleft$) and 512 
($\blacktriangledown$).
}
\label{fig3}
\end{figure}

We define a critical time for saturation $\tau_{\rm c}$ as the inverse 
of the frequency for which the value of the susceptibility is 95 \%\ of 
the asymptotic value for extremely long times. We studied the $T$ 
dependence of this critical time $\tau_{\rm c}$. For all temperatures
considered the values of $\tau_{\rm c}$ are extremely large, which is
a sign of the glassy nature of our systems. Since $\tau_{\rm c}$ is
long and close to the saturation of $\chi_0$, we expect our results to 
be rather accurate for this study. Fig.\ \ref{fig3} plots the logarithm
of $\tau_{\rm c}$ {\it vs.} $T^{-1/2}$ for four sizes of the standard 
model and for two sizes of the CIB model. The data are fitted quite well 
by straight lines indicating that a similar mechanism which gives rise to 
the $T^{-1/2}$ conductivity \cite{PO97} is also effective in the dielectric 
susceptibility. This should of course be expected because of the close 
connection between the two properties. The square root of the slope of 
each straight line yields a characteristic temperature $T_0$ which has 
been often associated with variable range hopping theory in a Coulomb 
gap. In that theory $T_0$ is given by:
\begin{equation}
T_0=\beta \frac{e^2}{4\pi\epsilon_0\, k_B a}
\end{equation}
where $\beta=2.8$ in three dimensions \cite{N84}. Such a theory does not 
take into account correlations, so comparing our results with the theory 
can assess the importance of many-body effects. The values of $\beta$ 
obtained from Fig.\ \ref{fig3} are $0.9\pm 0.2$ for the standard model 
and $0.9\pm 0.1$ for the CIB model, i.e., about three times less 
than in the one-particle theory. This indicates the importance of 
correlations. The results follow the same trend we found for two 
dimensional systems by a numerical simulation study of the conductivity 
\cite{PO97} where we also found $\beta$ systematically smaller than was 
predicted by the one-particle theory, and were able to identify specific 
many-body processes.

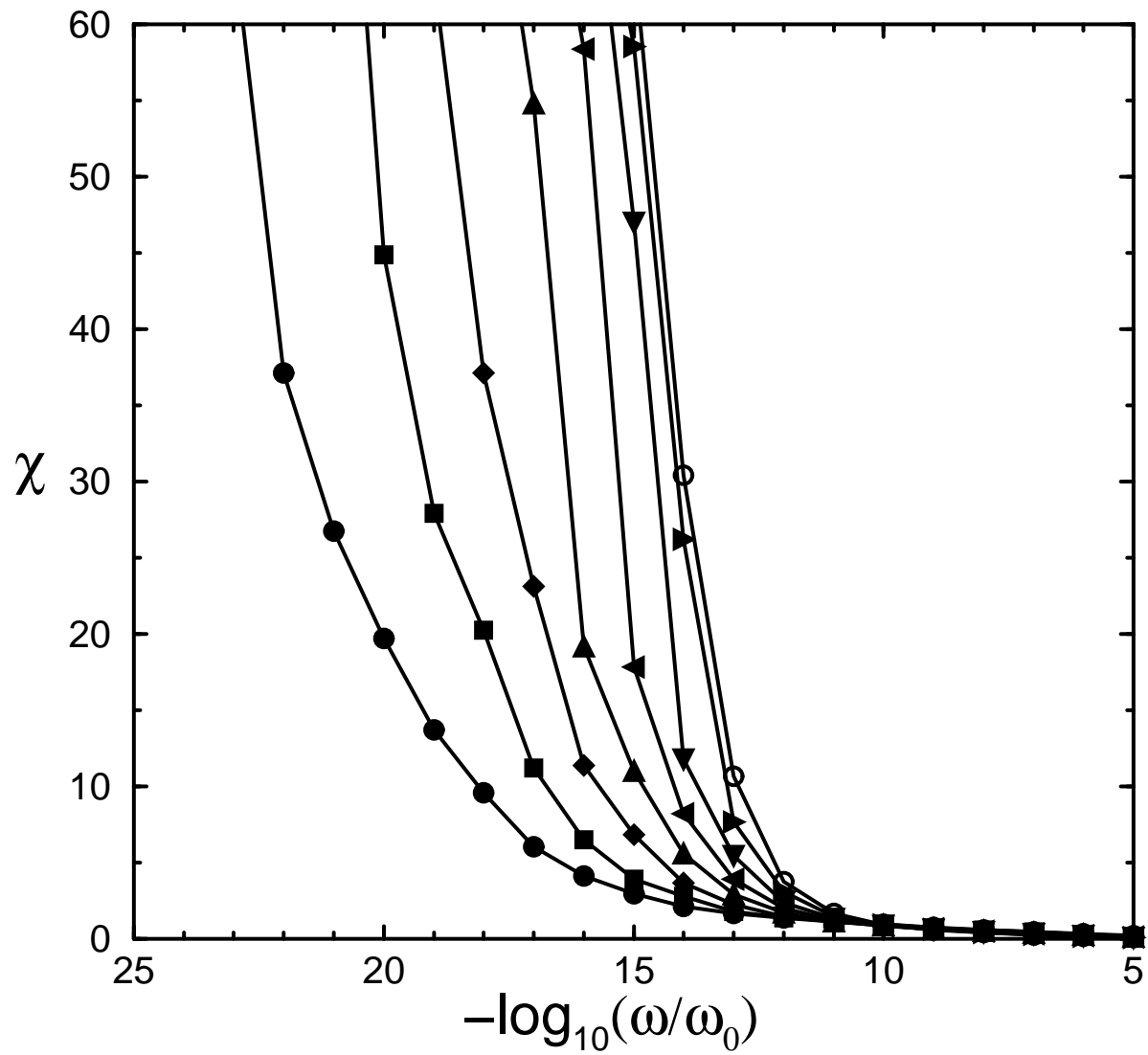
\begin{figure}
\caption{Dielectric susceptibility $\chi$ as a function of frequency 
for several temperatures, $T=0.006$ ($\bullet$), 0.008 ($\blacksquare$), 
0.01 ($\blacklozenge$), 0.012 ($\blacktriangle$), 0.014 
($\blacktriangleleft$), 0.016 ($\blacktriangledown$), 0.018 
($\blacktriangleright$), and 0.02 ($\circ$).The other parameters and 
the model considered are the same as in Fig.\ \ref{fig1}.
}
\label{fig4}
\end{figure}

To evaluate, at least approximately, the frequency behavior of the 
macroscopic susceptibility we calculated $\chi(\omega)$ using 
Eqs.\ (\ref{dipp}) and (\ref{fin}). The results are exhibited in 
Fig.\ \ref{fig4} where $\chi$ is plotted {\it vs.} frequency for several
temperatures. The data correspond to the data for $\chi_0$ of 
Fig.\ \ref{fig1}. Note that the susceptibility diverges in the static 
limit at all temperatures examined, but at a different rate at each 
temperature. Coulomb glasses will screen like a metal if where not for
their glassy nature, which produces typical times for the divergence of
the dielectric function much larger than measurement times.

\section{Conclusions}

We derived a microscopic expression for the dielectric susceptibility 
$\chi$ applicable to hopping systems including systems where interactions 
are important. The expression is particularly suitable for low frequencies. 
It corresponds to the expression used in classical electrodynamics. Si and 
Varma \cite{SV98} have recently study the same problem in the metallic limit
of the metal-insulator transition for two dimensional systems, and obtain
than the static compressibility vanishes at the transition. Some previous 
works \cite{DL84,GY93} used expression based on an analogy between spin and 
Coulomb glasses. We argues that these analogies cannot be extended to the 
susceptibility. The fundamental reason is that unlike in spin glasses the
susceptibility in the hopping systems arises from non-local processes.

The fluctuation--dissipation theorem tells us that the dielectric 
susceptibility is a function of the thermal fluctuations of the dipole 
moment of the system, instead of the fluctuations of the charge density,
result that one obtains when using analogy between Coulomb and Spin 
glasses. We calculate $\chi$ numerically for three--dimensional Coulomb 
glass systems as a function of temperature and frequency.  
We found that $\chi$ diverges as the frequency tends to zero. One has
to consider sizes larger than a critical one of approximately $N=200$ 
for the CIB model and for the standard model with $W=3$. 

The logarithm of the critical time for saturation varies proportionally
to $T^{-1/2}$, the same dependence as in variable range hopping. 
The characteristic temperature for this dependence is 
approximately equal to 0.9, a factor of three smaller than the 
theoretical predictions for the equivalent constant appearing in 
variable range hopping.

\acknowledgments

We acknowledge financial support from the DGES project number PB96-1118,
SMWK, and DFG (SFB 393). A great part of this work was performed during 
A.~D.-S.'s visit at the IFW Dresden; A.~D.-S.\ thanks the IFW for its 
hospitality.

\end{document}